# WatSen: Design and testing of a prototype mid-IR spectrometer and microscope package for Mars exploration

S. D. Wolters[a], J. K. Hagene[b], A. T. Sund[c], A. Bohman[b], W. Guthery[a],

B. T. Sund[c], A. Hagermann[a], T. Tomkinson[a,d], J. Romstedt[e], G. H. Morgan[a], M. M. Grady[a,f]

[a]Department of Physical Sciences, The Open University, Milton Keynes, MK7 6AA, UK [1]

[b]Norsk Elektro Optikk AS, Solheimveien 62 A, N-1473 Lørenskog, Norway

[c]NavSys AS, Tevlingveien 23, 1081 Oslo, Norway

[d]Scottish Universities Environmental Research Centre, East Kilbride, Scotland, G75 0QF, UK

[e]ESA/ESTEC Keplerlaan 1, 2200 Noordwijk, The Netherlands

[f]Department of Mineralogy, The Natural History Museum, Cromwell Road, London SW7 5BD, UK

**Address for Correspondence:** Dr S. D. Wolters (stephen.wolters@jpl.nasa.gov; s.d.wolters@open.ac.uk)

---

[1] Currently employed by the Jet Propulsion Laboratory, California Institute of Technology. Work contributed to this paper was not done in the author's capacity as an employee of the Jet Propulsion Laboratory, California Institute of Technology

**Abstract**

We have designed and built a compact breadboard prototype instrument called WatSen: a combined ATR mid-IR spectrometer, fixed-focus microscope, and humidity sensor. The instrument package is enclosed in a rugged cylindrical casing only 26mm in diameter. The functionality, reliability and performance of the instrument was tested in an environment chamber set up to resemble martian surface conditions. The effective wavelength range of the spectrometer is 6.2 – 10.3 μm with a resolution Δλ/λ = 0.015. This allows detection of silicates and carbonates, including an indication of the presence of water (ice). Spectra of clusters of grains < 1mm across were acquired that are comparable with spectra of the same material obtained using a commercial system. The microscope focuses through the diamond ATR crystal. Colour images of the grains being spectroscopically analysed are obtainable with a resolution of ~ 20 μm.

# 1   Introduction

The occurrence on Mars of large volumes of subsurface water ice at depth was confirmed by the MARSIS instrument onboard ESA Mars Express (Plaut et al., 2007). In situ studies by instruments onboard the Phoenix lander observed subliming ice within the top few cm of regolith (Rennó et al., 2009). Liquid water, however, has not been detected, even though there are abundant signs of its presence throughout Mars' history.  Surface measurements from Mössbauer spectrometers on the Spirit and Opportunity rovers indicate the presence of haematite, jarosite and gypsum (Klingelhöfer et al., 2004; Squyres et al., 2012); the occurrence of these minerals on Earth is typically dependent on the action of water. In the shallow subsurface (< 1m), measurements of neutrons and gamma-rays by the Gamma Ray Spectrometer on the Mars Odyssey spacecraft (Boynton et al., 2002; Mitrofanov et al 2002, 2003) have found that water ice permafrost is present in a subsurface layer poleward of +/- 40 degree latitude, covered by a hydrogen-poor surface layer of varying thickness.  Further evidence for Mars' fluvial past is found within martian meteorites. Although typically igneous in origin, small quantities (< 1% by weight) of secondary alteration components (carbonates, sulphates and clay minerals) are present, particularly in the nakhlite subgroup (Bridges et al., 2001). These secondary minerals are produced by an aqueous medium, and are concentrated along cracks within the meteorites (Bridges and Grady, 2000). Isotopic measurements show that the carbonates are martian, and not terrestrial contaminants (Carr et al., 1985; Wright et al., 1992).

In 2005, the European Space Agency (ESA) issued a Statement of Work (AO 4880) defining the requirements for an instrument package that would be able to: (1) detect thin layers of water (or ice) surrounding particles below the surface of Mars and (2) measure the humidity at and below the surface of Mars. The instrument package (with dimensions specified by ESA) was designed to be a combined miniaturized attenuated total reflectance (ATR) spectrometer and humidity sensor that could be integrated into a cylindrical compartment. The resource requirements were derived from a carrier system, called the Mole, that is capable of penetrating a regolith layer down to a depth of 5 m (Gelmi et al. 2007). A suite of appropriate sensors delivered by the Mole into the regolith would make the first observations of Mars' non-oxidised subsurface. After initial design work, the scope of the project was expanded to include a fixed-focus microscope into the package. The resulting combination is WatSen (for Water Sensor), an integrated package of three instruments: an infrared (IR) spectrometer, a microscope and a humidity sensor. WatSen was designed and built by Norsk Elektro Optikk (Norway), responsible for optics and mechanical design, and NavSys (Norway), responsible for electronics, software, and the humidity sensor. Project management and testing were undertaken by the Open University (UK).

The main objective of the WatSen instrument package would be to detect water adsorbed on the surface of soil grains (Grady, 2006). Liquid water will not be stable within the uppermost soil layers, the zone of sublimation. Ambient pressure is such that liquid water would evaporate instantly and thin layers of ice sublime. The depth of the zone of sublimation has yet to be determined accurately. Below the zone of sublimation, depending on the porosity of the regolith, water ice is likely to be stable. Water vapor could also exist, i.e. the atmosphere within pore spaces could have high humidity. In order to interpret humidity data fully, a measure of the porosity of the substrate is required. This could come from permittivity data acquired by complementary instrumentation, e.g. the HP$^3$ (Heat Flow and Physical Processes Package; Grott et al., 2009).

Secondary goals of WatSen would be to determine the mineralogy and mineral chemistry of soils, anticipating that during deployment, the WatSen package could be employed to determine any changes in mineralogy with depth, and also how grain size, grain morphology and porosity might change with depth. The main rock type on Mars' surface is basalt, which is composed of anhydrous silicate minerals, mainly pyroxene and plagioclase, plus olivine. These minerals are primary, i.e. were produced at depth within a magma chamber, and crystallised during the cooling of a high temperature melt. A different suite of minerals, including carbonates and phyllosilicates, is formed at lower temperatures by aqueous alteration and aqueous deposition. Carbonates are of interest because they require water and carbon dioxide for their formation, both of which are known to exist on past and present Mars. They are also commonly associated with organics on Earth. Phyllosilicates have been detected by OMEGA/ Mars Express (Bibring et al. 2005; Poulet et al. 2005), and were also found by the MER rovers (Glotch et al., 2006). Carbonates have been found as a minor phase within martian meteorites (< 1% by volume) and within martian dust (~ 2-5 weight %, dominated by magnesite), using the Thermal Emission Spectrometer on board the Mars Global Surveyor spacecraft (Bandfield et al., 2003). The general lack of large scale deposits is puzzling given other evidence for aqueous alteration and has been ascribed to the carbonates being destroyed by acidic aqueous activity (Fairén et al, 2004). However, Ehlmann et al (2008), using CRISM/Mars Reconnaissance Orbiter, discovered a rock layer in the Nili Fossae region containing magnesium carbonate; found in an area associated with clays, this suggests that acidic weathering did not dominate in all regions. In situ discoveries of calcite were made by the Phoenix lander (Boynton et al., 2009) and Mg-Fe carbonates were found within outcrops by the Spirit rover (Morris et al., 2010), suggesting subsurface hydrothermal origins. It was anticipated that by selecting an appropriate wavelength range for the detector, WatSen could also investigate the silicate mineralogy of the regolith, and identify the presence of secondary minerals including phyllosilicates and carbonates.

## 2 Instrument Design

### 2.1 Design Specifications

Grady (2006) outlined the required scientific specifications from ESA for WatSen to be considered viable for attachment to a mole in a future planetary exploration mission. The completed compact breadboard model surpasses ESA's original design specifications in terms of size, mass and power budgets (Table 1 and Figure 1).

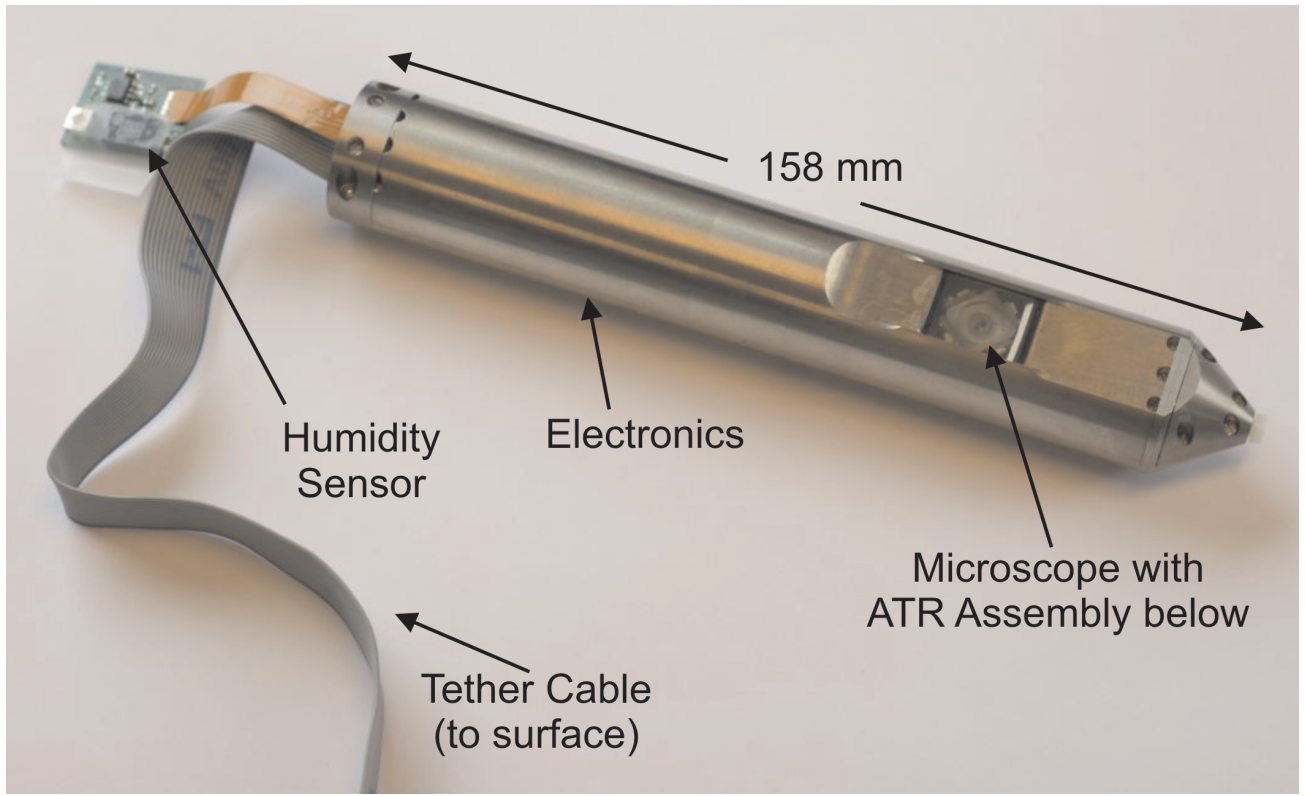

**Figure 1:** Completed breadboard model of the WatSen instrument package (April 2009).

**Table 1:** WatSen - specified compared to achieved parameters

| Parameter | Specification | Achieved |
|---|---|---|
| Power | 3 W | < 2.8 (peak); < 2.1 (average) |
| Mass | 300 g | 190 g |
| Shape | Cylindrical; Ø 26 mm | Cylindrical; Ø 26 mm |
| Length | 158 mm | 158 mm |
| Temperature range | 223 – 323 K | 223 - 298 K |
| Pressure | <10 mbar | 5 mbar |
| Spectrometer wavelength range | | 6.2 – 10.3 μm |
| Spectrometer resolution (Δλ/λ) | | 0.015 |
| Microscope field of view | | 14.5 x 8.5mm |
| Microscope resolution | | 18 μm at centre |

The main constraint on the physical size of the package was that it had to fit within the mole subsystem (Figure 2) intended as a deployment mechanism for sensor packages. The mole was designed under ESA contract by DLR, Berlin (Gelmi et al., 2007). The sensor package would be carried by a trailed mole, and so had to be compatible in size with this mechanism, giving a requirement for WatSen to fit within a cylinder 26 mm in diameter and 158 mm in length. The design also had to be sufficiently robust so that WatSen could survive deployment underground if the mole had to hammer its way into the regolith.

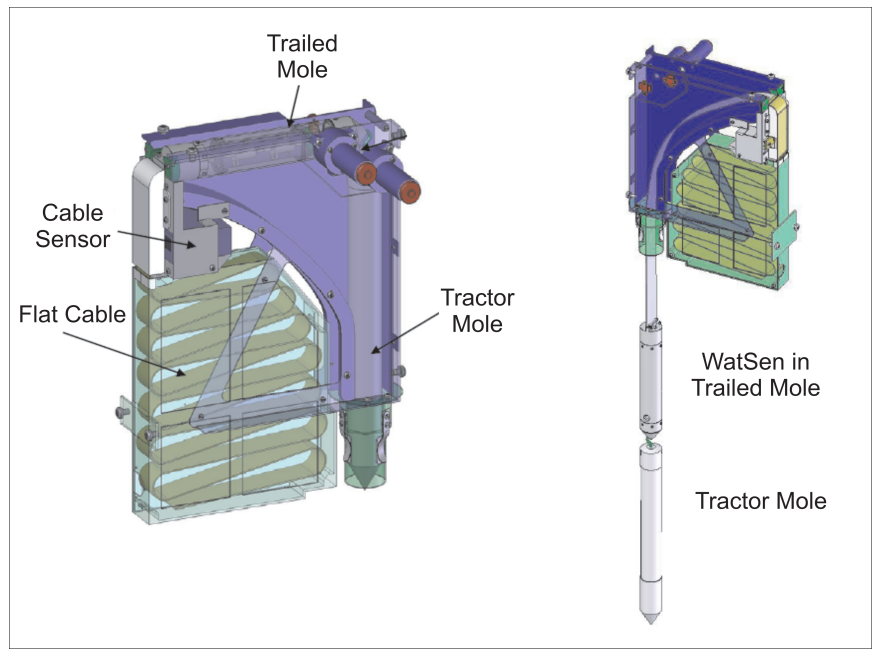

**Figure 2:** Schematic cartoon of the tethered mole into which the WatSen package was designed to fit (adapted from Gelmi et al., 2007).

The second limit on the design of WatSen was the wavelength range over which the detector would operate. This was constrained by the detector assembly, which could not be large or complex. Trade-offs between the size of the detector array, the strength of the water feature and the presence of other features in mineral spectra led to selection of a desired mid-IR wavelength range of 5.5-11 µm. Figure 3 shows the features that occur in this range in a selection of species appropriate for investigation of Mars' surface. This wavelength range is a trade off between technical boundaries and scientific capability; liquid water absorption is strongest at 2.9 µm, but there is also a strong and distinct feature at 6 µm (Roush et al., 1991). Anhydrous silicate minerals have absorption features in the mid-IR at wavelengths ~ 9 - 12 µm (Roush et al., 1991), whilst different carbonate signatures occur between 6.3 - 7.4 µm. Furthermore, hydrated minerals, such as clays, present the combined features of water and silicates, and so can also be identified within the designated wavelength range. Ultimately, testing of the completed device measured low signal-to-noise at the edges of the wavelength window, such that the effective range is found to be 6.2-10.3 µm. We would still be able to distinguish between a pure sample of pyroxene, olivine or plagioclase, and identify carbonates. However, the scope of the present study only employed one rock sample, rather than a range which would define WatSen's discriminatory powers more precisely (c.f. Figure 11).

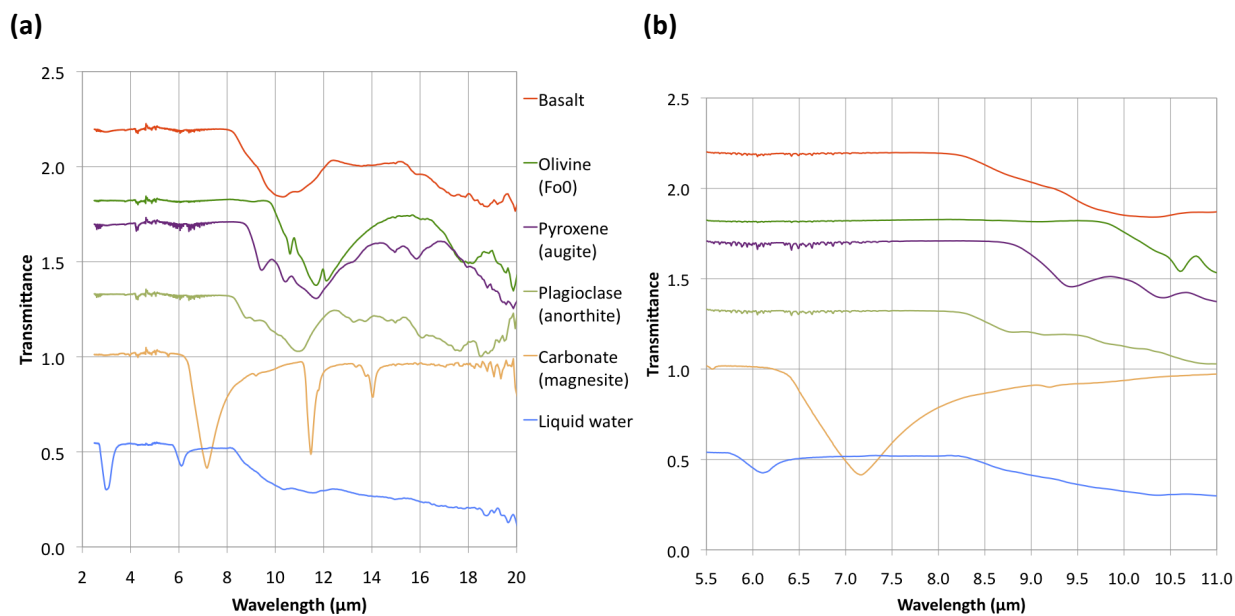

**Figure 3:** Features of interest in the Mid-IR spectra from water and a range of minerals. (a) Whole spectrum; (b) spectral range selected for WatSen operation. Spectra are slightly offset from each other for clarity.

## 2.2 ATR spectrometer

Infrared spectroscopy is a rapid and non-destructive technique that detects energy changes during stretching, bending and vibration of intra-molecular bonds, and is an effective method for determining the presence or absence of water. It can be employed in either transmission or reflection mode; in the former case, the thickness of a specimen is an important parameter for measurement, but not in the latter. Hence reflectance spectroscopy is the method selected for planetary exploration instruments which rely on small scale contact analysis. However, spectral reflectance features are almost always dependent on surface properties (flat or uneven, rough or smooth, etc.), so the signal generated must be from a representative grain surface. This is best achieved by placing the sensor in direct contact with the surface of the grain; the most appropriate sensor type to achieve this measurement is an Attenuated Total Reflection (ATR) sensor.

ATR spectroscopy is a powerful technique for studying the absorption IR spectra of a variety of materials. It is based on the phenomenon of total internal reflection, occurring when electromagnetic waves in a transparent dielectric medium impinge onto the surface of an absorbing medium with a lower index of refraction, and with an angle of incidence above the critical angle. An evanescent wave is excited which propagates a certain distance (penetration depth) into the sample (Figure 4). Some of the energy of the evanescent wave is absorbed by the sample at particular wavelengths and this affects the signal received by the detector. The ATR sensor utilises the effect that the reflectance properties of a mineral grain are altered when the grain is coated with a thin layer of water.

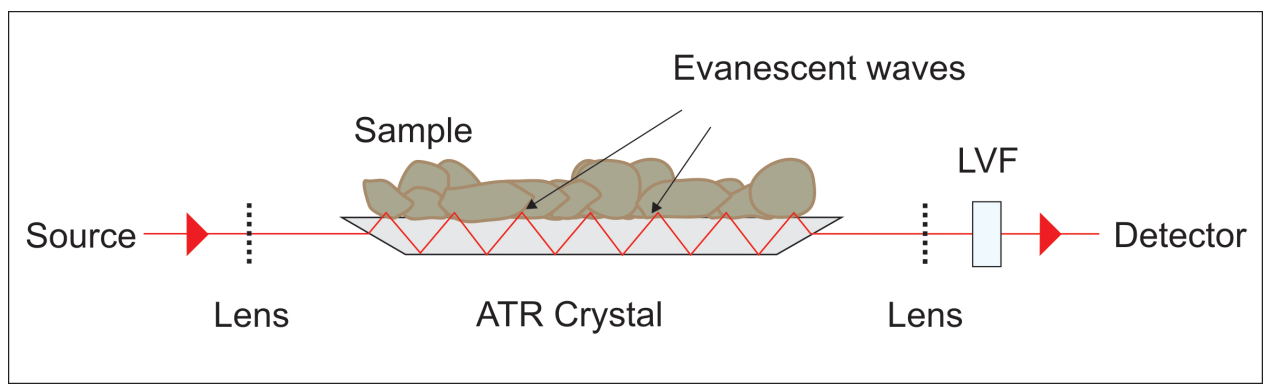

**Figure 4:** Iillustration of the light path through an ATR crystal and the formation of an evanescent wave. LVF – Linear variable optical bandpass filter.

The ATR spectrometer (Figure 5) comprises of a thermal infrared source, a diamond ATR crystal, and a pyroelectric linear sensor array. Coupled with a linear variable optical bandpass filter, the sensor array provides the wavelength separation and detection capability essential to the function of a spectrometer. The ATR spectrometer light source is a MIRL 17-900 blackbody thermal emitter. The emitter surface is a thermoresistive film of conducting amorphous carbon heated electrically to temperatures up to 750 °C. The microstructured film emits a blackbody spectrum with an emissivity of 0.8. The radiation is collimated into a beam by a system of two anti-reflection-coated lenses in front of the infrared source. The beam then passes through an aperture that forms a well confined beam before the beam passes through the diamond. The aperture is blackened with a dark optical coating to reduce stray light.

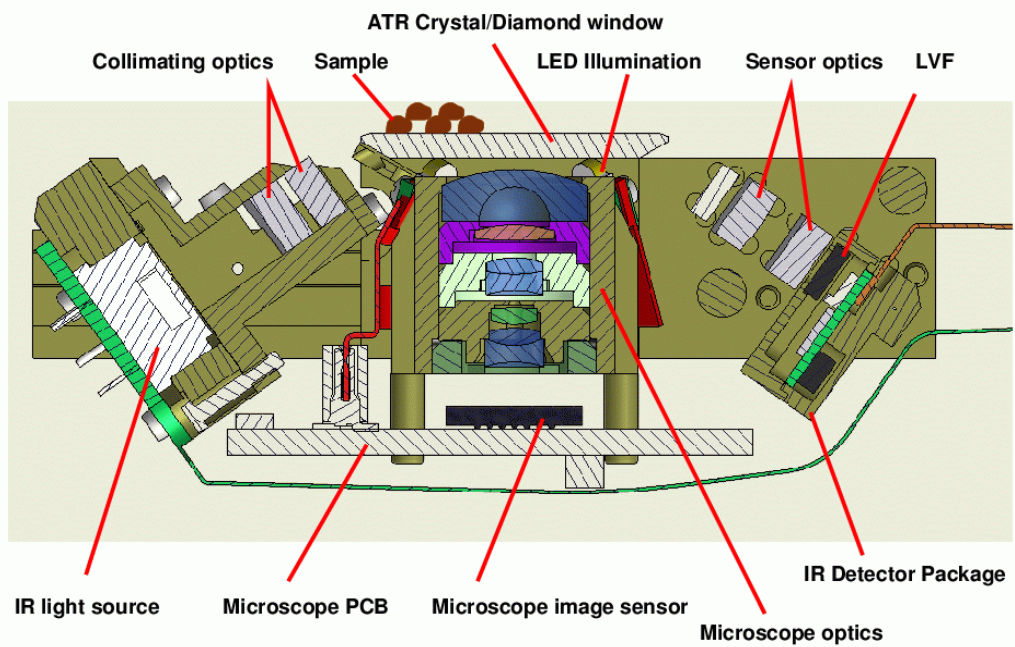

**Figure 5:** ATR spectrometer and microscope design. The optical system for the ATR sensor contains two lens assemblies, a collimator system in front of the source and a focusing lens assembly in front of the detector. The microscope lenses are made from antireflection-coated germanium.

The ATR crystal could be made from any material that is transparent to radiation at the required wavelengths. Possibilities for WatSen included Ge and $MgF_2$, but diamond was chosen because it is durable and scratch-resistant, as well as relatively transparent to radiation at the required wavelength range. The diamond in WatSen is a CVD (Chemical Vapour Deposition) grown polycrystalline synthetic diamond, 19 × 13.5 × 1.5 mm in size. The length of the crystal was set such that there would be 7 internal reflections of the beam between entry and exit (Figure 4). The 1.5-mm thickness was the maximum available from the manufacturer (Element Six). Limitations of the manufacturing process cause such thick diamond windows to have a polycrystalline structure, which in the visible wavelength range manifests itself as numerous scattered dark specks in the diamond. This is mainly an issue for the microscope; in the infrared, the impact on light propagation is greatly reduced.

The infrared beam is coupled in and out of the crystal through angled facets which are coated to avoid high reflection losses. Inside the diamond, the transmitted infrared beam undergoes multiple total internal reflections with an incident angle of 40°. As the radiation exits the diamond, the beam passes through a focusing lens assembly of two coated germanium lenses.

The lens system directs and focuses the beam through a linear variable bandpass filter (LVF). This transmits a narrow spectral band of the incident light onto the detector, which is an infrared sensor (DIAS Infrared 128LTI SP1.0) in a custom-designed package. The sensor is an uncooled pyroelectric 128-element linear array with integrated CMOS multiplexer including amplifiers. The dimensions of the responsive elements are 90 × 1000 µm with a pitch of 100 µm (array size 1 mm × 12.8 mm). The pyroelectric sensor elements respond to temperature changes from alternating heating and cooling of the sensor elements, resulting in chopped radiation incident on the sensor (some light is still radiated in the off periods). The effective resolution of the sensor is in general less than 128 spatial channels (corresponding to 40 nm spectral resolution) as a result of thermal diffusion into neighbouring pixels at low chopping frequencies. 10 Hz chopping frequency was selected as a reasonable compromise between response and resolution.

The multiple (7) reflection ATR crystal design has bands that are sensitive and bands that are more or less blind to the sample (Figure 6). Spectral sensitivity is also different depending on where on the window the sample is located. This is partly a consequence of dispersion of the parallel beam through the diamond, which, in essence, splits the beam by wavelength (somewhat like a prism). The dispersed radiation falls onto the LVF, which filters the light such that radiation with wavelength 5.5 µm will be transmitted to one end of the detector array, and that with wavelength 11 µm to the other end. Figure 6 illustrates how this leads to different spectral sensitivity for different spatial positions on the window.

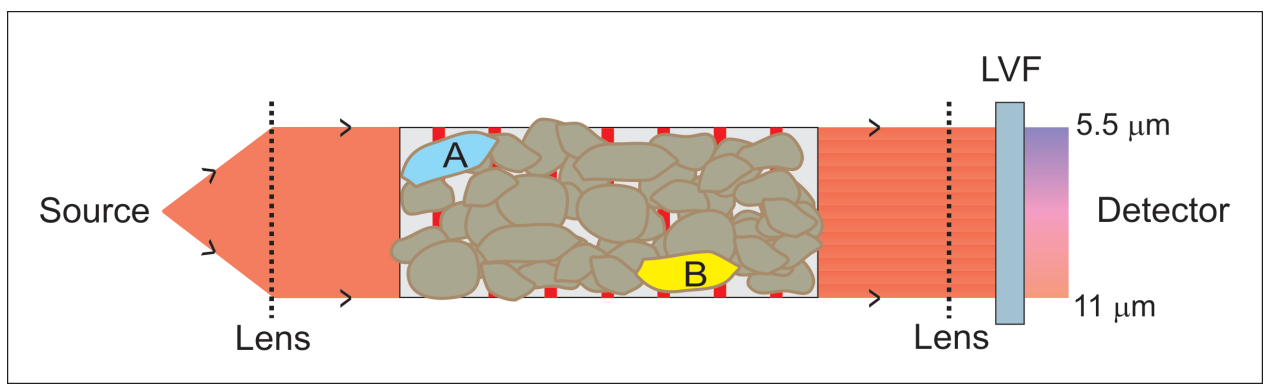

**Figure 6:** Cartoon (not to scale) of a plane view of the light path through the ATR optics, to illustrate dispersion of radiation through the ATR crystal, and the differing results that arise, depending on the position of material on the crystals. The red stripes across the crystal indicate the ATR 'bounce' positions, where sample and radiation interact. For the two different materials on the ATR diamond window, sample A will contribute to the acquired spectrum from around 5.5 µm to around 7 µm. The spectral properties of sample A in the range 7-11 μm are lost. Sample B contributes to the spectrum from around 9 μm to 11 μm while all spectral features of sample B in the range 5.5-9 μm are lost.

The resulting spectra from samples A and B in Figure 6 are only partial; if a spectrum across the entire wavelength range is required to discriminate sample A and B from each other or from other grains, the sample must cover the entire width of the window. Thus this measurement concept works well if a sample is thoroughly mixed and can be distributed over the ATR crystal window.

## 2.3 Microscope

Visual information about the distribution of particles on the diamond window is important, since the strength of the absorption features seen in the spectra depends critically on the contact between sample and the window surface. A modification of the original spectrometer design incorporated a microscope that would be able to view the sample as it impinged on the ATR spectrometer window. The design placed an additional constraint on the window: because the ATR crystal must also act as an optical window through which an image would be viewed, it had to be transparent to visible as well as IR radiation. Since the microscope must be located behind the ATR crystal, the ATR spectrometer source and detector modules had to be sufficiently separated to accommodate the microscope.

The microscope is a miniature fixed-focus camera, which images a sample located in front of the diamond ATR window. Four white LEDs arranged around the front end of the microscope provide the necessary illumination. The lens system (Figure 5) has a wide field of view that covers nearly the full width of the outer diamond surface, with a depth of field of around 1 mm. The field of view comes at the expense of strong distortion; the magnification decreases by more than 50% towards the edge of the image ("barrel"-type distortion). However, it was demonstrated during development that this distortion can be corrected by digital post-processing of the image. The lens system images the sample onto a 1280 × 1024 pixels 1/3 inch CMOS Bayer colour filter array mounted on a PCB. The resolution of the image sensor exceeds that of the lens system, which is limited by diffraction. In the centre of the image the effective resolution of the microscope is 18 μm (32 μm at edge). The performance of the microscope is given in Table 2:

**Table 2:** WatSen microscope performance

| Parameter | Centre | Edge |
|---|---|---|
| Field of view, mm | 14.5 x 8.5 | |
| Magnification | 0.5x | 0.33x |
| Spatial frequency (image) lp/mm | 110 | 95 |
| Spatial frequency (object) lp/mm | 55 | 31 |
| Resolution (object) (µm) | 18 | 32 |
| Contrast, % | 30 | 30 |

## 2.4 Humidity Sensor

The final WatSen breadboard design contained a capacitive humidity sensor. We chose a commercial capacitive sensor, the Humirel HTS2030 sensor with an integrated NTC (Negative Temperature Coefficient) thermistor which was mounted on the flex cable at the rear of WatSen. Additionally a Sensirion SHT15 was used for monitoring during the environmental chamber tests.

## 2.5 Mechanical Design

The body of WatSen consists of two half cylinder shells that encapsulate the payload compartment (Figure 7). At the front is a conical nose section with an electrical feedthrough, intended for connection to a tractor mole. Another electrical feedthrough at the rear is attached to a flat ribbon cable, leading to the control electronics. In a flight-ready instrument, this would connect to the tether. The atmospheric humidity sensor is attached to the ribbon cable.

Most of the mechanical parts were fabricated from titanium, selected to provide realistic instrumental properties in terms of thermal expansion and conductivity. The internal volume of WatSen was hermetically sealed after being baked out and filled with dry nitrogen. Silicone elastomer gaskets provide sealing between titanium components of the outer body. The electrical feedthroughs in the nose and rear end are potted with epoxy. The diamond window is glued into the titanium shell with a silicone-based encapsulant.

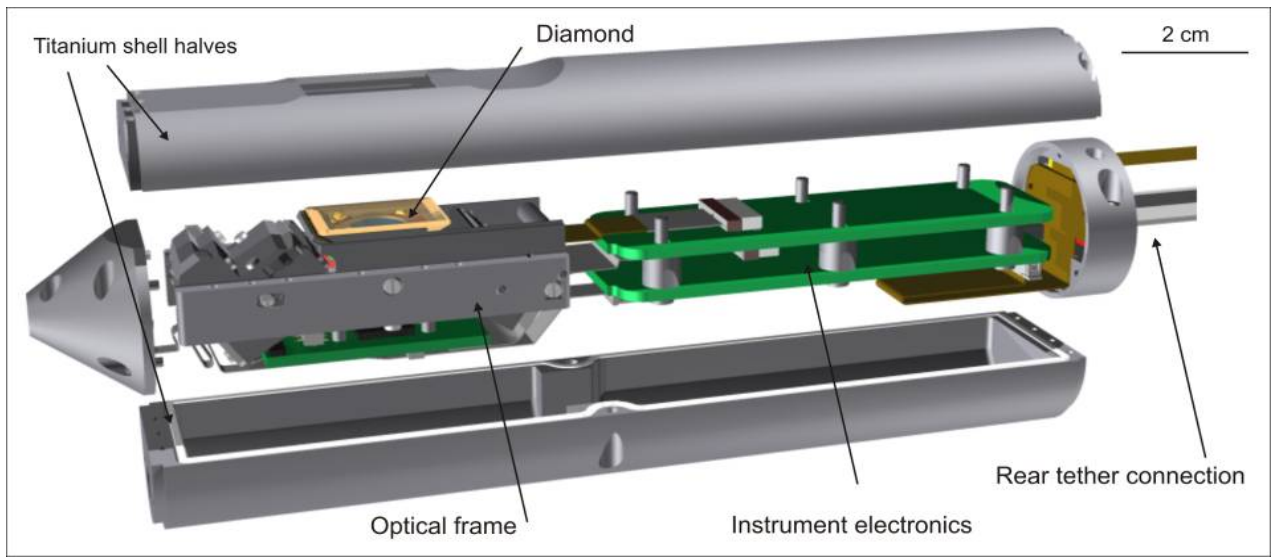

**Figure 7:** An exploded view of the mechanical layout of WatSen. The optical frame contains the ATR infrared source and detector modules as well as the microscope. Flexible PCBs connect the instrument electronics to the infrared detector, infrared source, and image sensor PCB as well as the nose feedthrough and rear tether.

**2.6 Electronics**

The electronics are housed in the rear half of the WatSen shell (Figure 7). The small volume available for the electronics resulted in a compact design of 2 multilayer PCBs, each with components on both sides. The electronics package comprised a power supply, and processing modules for the microscope image sensor, humidity sensor and FGPA.

The two main circuit boards are supplemented by the Microscope Board, carrying the image sensor, and the Humidity Sensor Board, located on the tether cable. Because of the low atmospheric pressure requirements, the electronics had to be conduction-cooled. This was realized by using six attachment points for each main circuit board (Main Module and Power Supply Module), solid copper inner layers on the circuit boards and suitable thermal conduction paths to the titanium enclosure.

The electronics design is based on a Field Programmable Gate Array (FPGA) which made it possible to keep the electronics within the specified mechanical envelope and limited the parts count. Low-voltage operation and the high-efficiency power regulation and switching system of the Power Module also contributed significantly to the low operating power. The FPGA circuitry supports:

- Power regulation and switching
- High-resolution A/D conversion
- Low-level, low-noise signal amplification
- Memory (image data buffer)
- Interface devices
- IR source
- Sensors (IR and image sensor)
- Temperature sensors

Additional features in the electronic design included:

- High-performance 32-bit RISC microprocessor for signal, image processing, data handling and system management
- High-speed telemetry interface (256 kbit/s vs. 19.2 kbit/s requirement)
- SpaceWire I/F compatible for optional megabit/s range data transfer rate
- Integrated heaters supporting operation at extremely low temperatures
- Unregulated, single supply 10-32 VDC operation
- JTAG (IEEE 1149.1) test bus and programming I/F

Thermal management of the overall system was also controlled within the electronics package. Whilst the IR source heats itself, the analogue detector amplifiers, the detector, the microscope and illuminations LEDs, as well as the general electronics could be heated to -55 °C using heat films on the printed circuit boards.

## 3. Performance tests

The test campaign had the aim of showing that the instrument works within a range of environment conditions, i.e. that the instrument had the required resolution and high enough S/N within a large enough wavelength range that we could detect water, silicate features, and carbonates. Also, we tried different percentages of water and carbonates to get a rough handle on sensitivity.

The main test material for WatSen was basalt (labelled 'Rock D'), several kilograms of a which were crushed and then sieved into two fractions; coarse grains 1-2 mm and fine < 1 mm in diameter. The sample was characterised using a Thermo Nicolet Nexus FTIR spectrometer (FTIR TNNS) coupled with a Continuum IR microscope with a spectral resolution ~ 1cm$^{-1}$ at mid-range; it is a typical medium-grained basalt, i.e. composed of pyroxene, plagioclase and olivine (Figure 8).

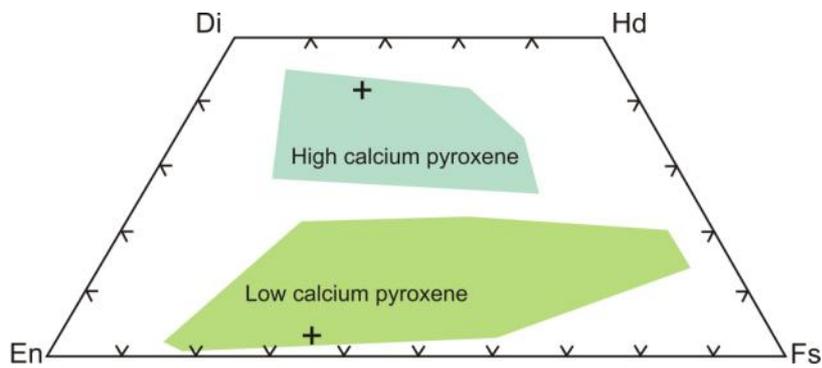

**Figure 8:** Pyroxene quadrilateral showing the range of high and low calcium pyroxene compositions exhibited by martian meteorites. The crosses are the compositions of pyroxenes in rock D.

Testing of the WatSen package took place using an environmental chamber that could be evacuated, flooded with different gas mixtures and operated over a range of temperatures from 77 K to 373 K, coupled to a baseplate cooled with liquid nitrogen. WatSen was placed in a Test Vessel Assembly (TVA), which could be filled with powdered rock, to simulate WatSen's deployment buried in a granular near-subsurface (Figure 9). Cooling for the 223K tests would take about two days, during which WatSen would be continuously exposed to wet and then icy crushed rock, and then be functional and performance tested. A sequence of tests was carried out (Figure 10), designed so that WatSen's performance could be assessed under different environmental conditions, including low pressure and temperatures similar to those at the martian surface. Background tests, without a sample in the TVA, were performed for each set of conditions, to provide master background spectra for ratio with the raw spectra acquired. For simplicity the chamber was filled with nitrogen. The experiments were conducted by using dry basalt samples and also sample material containing different amounts of water. The standard procedure in each test was to take one test spectrum of 0.1 s exposure time (i.e. one integration), followed by the first microscope image, followed by 5 spectra, each of 100 s exposure time (i.e. 1000 integrations summed). Four batches of 5 spectra are taken, and between each batch another microscope image taken. All spectra given in the figures will be the median spectrum of (4 batches) × (5 spectra in each batch) ratioed with the appropriate master background spectrum.

**Figure 9:** (left) interior view of environmental chamber and TVA; (right) schematic plan of environmental chamber.

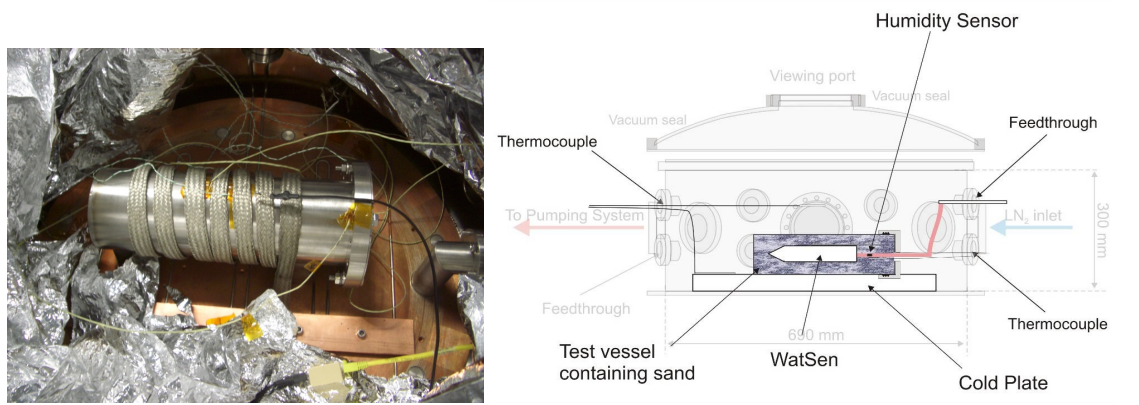

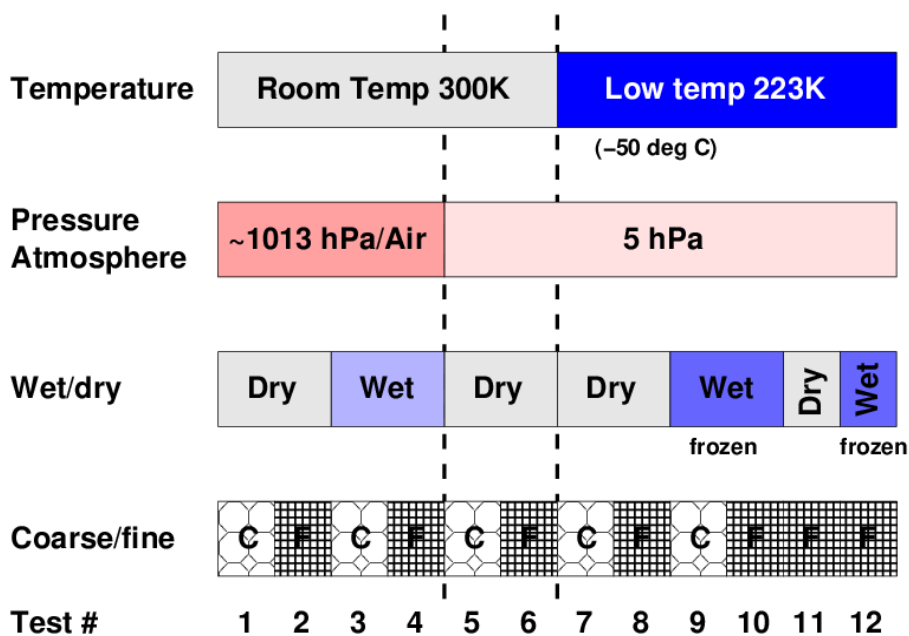

**Figure 10:** Schematic illustration of the test programme. Variation of temperature, pressure and moisture content of the samples as well as grain size (coarse or fine) are shown.

In testing, the Humirel humidity sensor gave appropriate (but uncalibrated) measurements of relative humidity before unfortunately failing. The readings showed that the sensor is sensitive to 1% weight by water and that a humidity sensor is potentially a useful tool to detect water where the ATR spectrometer is insufficiently sensitive. The environment chamber Sensirion sensor was operated successfully during all the tests, indicating that relative humidity would in principle be measurable under martian surface conditions.

A comparison of the room temperature and pressure basalt tests to those taken of a similar mixture using the commercial FTIR TNNS system is given in Fig. 11. We see that the data from coarsely-powdered basalt are comparable to those obtained using the FTIR, whilst that of finely-powdered basalt is much sharper and has more crisply-defined features. The signature of water is apparent in both of the runs where 5 wt% water was present, with a clear feature from water at around 6.5 μm. Between 6.2 and 10.3 μm the spectral resolution exceeds Δλ/λ = 0.015, with S/N sufficient to identify the broad water and silicate features (>50).

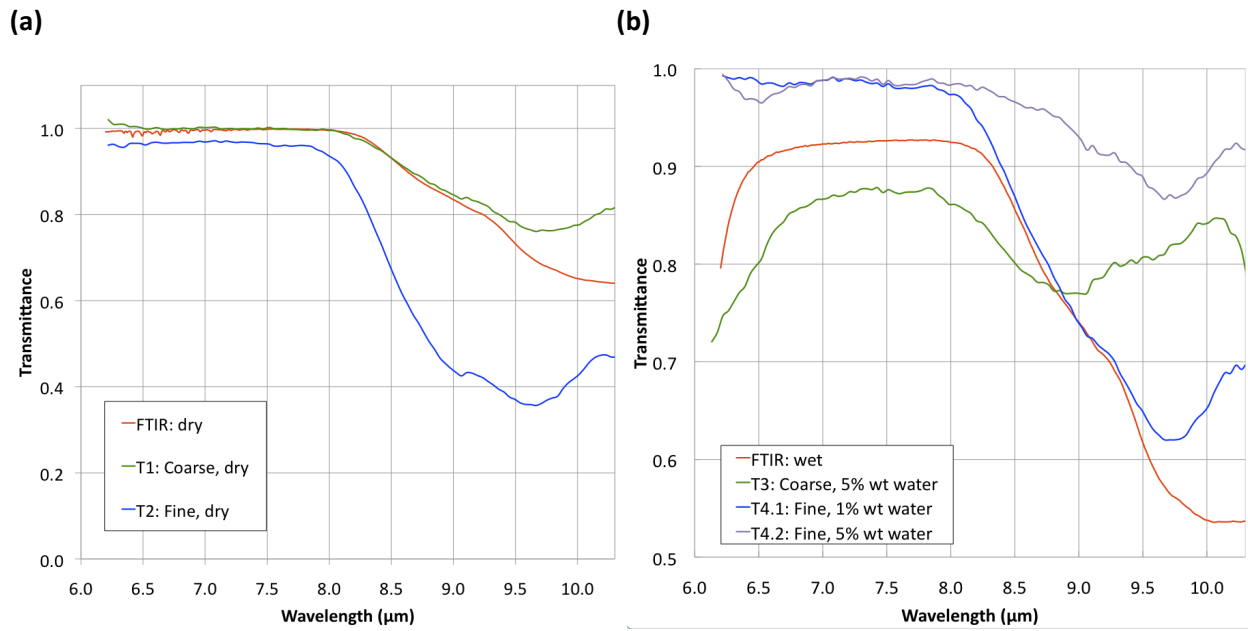

**Figure 11**: Room temperature test results. (a) Dry basalt. FTIR (red) is the spectrum of Rock D obtained using the commercial ATR system; (b) wet basalt: T3 (green) is the spectrum of coarsely-powdered basalt mixed with 5 wt% water; T4.1 (dark blue) is the spectrum of finely-powdered basalt mixed with 1 wt% water; T4.2 (light blue) is the spectrum of finely-powdered basalt mixed with 5 wt% water.

For the low temperature, wet tests, when ice forms on the surface of the ATR spectrometer there is a bigger divergence (Figure 12). For dry basalt, there is little difference in the results from room temperature and low temperature - the spectra have the same shape and exhibit the same features. The biggest effect on dry basalt seems to be in the difference between spectra acquired at atmospheric and low pressure. The wet spectra are noticeably different from their dry counterparts, with a very broad water feature present at the lowest wavelengths. The silicate feature at higher wavelengths has also broadened.

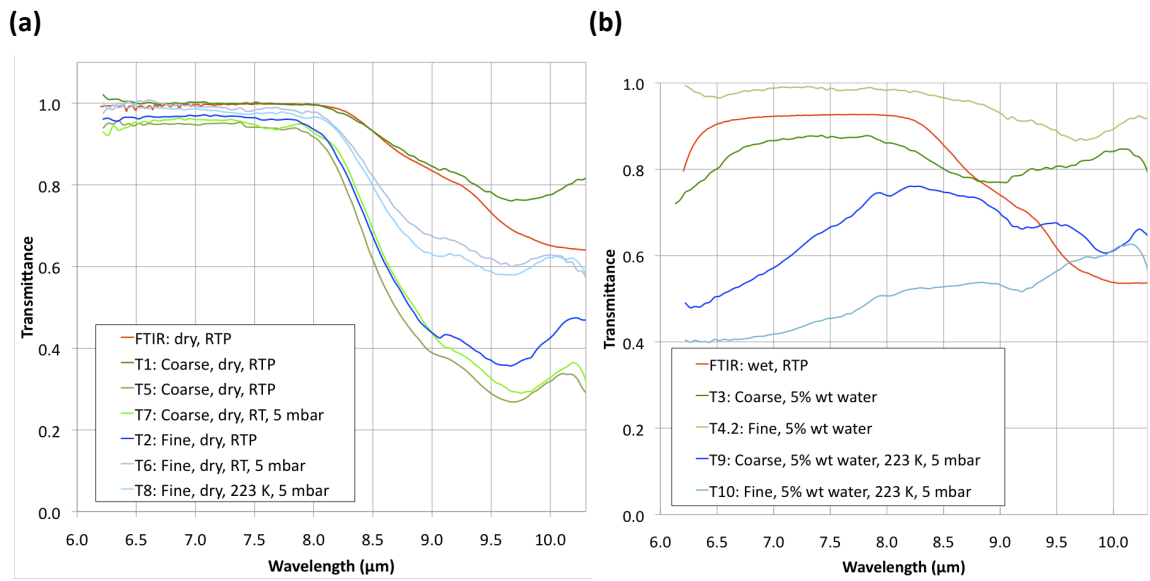

**Figure 12:** Low pressure/temperature test results. (a) Dry basalt, (b) Wet basalt. The room temperature and pressure results and the commercial FTIR TNNS are included on the figures for comparison.

The changes with temperature are illustrated by looking at individual batches for one low temperature wet test (Figure 13). The water feature is very broad and dominates the whole spectrum. There is a trend of increasing transmission (i.e., less water) with time. The soil temperature is increasing between batches, WatSen's surface temperature is rising and the pressure is rising fairly rapidly (Figure 13a). We speculate that as WatSen warms, water evaporates from the window, causing the depth of the feature to decrease. This can be seen in example microscope images from the test, which show progressively less ice covering the window for later batches. This is a good example of how the microscope was useful for interpreting the spectra during testing. Grains smaller than 1mm can be discerned, and the microscope was shown during functional testing to have the capabilities given in Table 2.

(a)

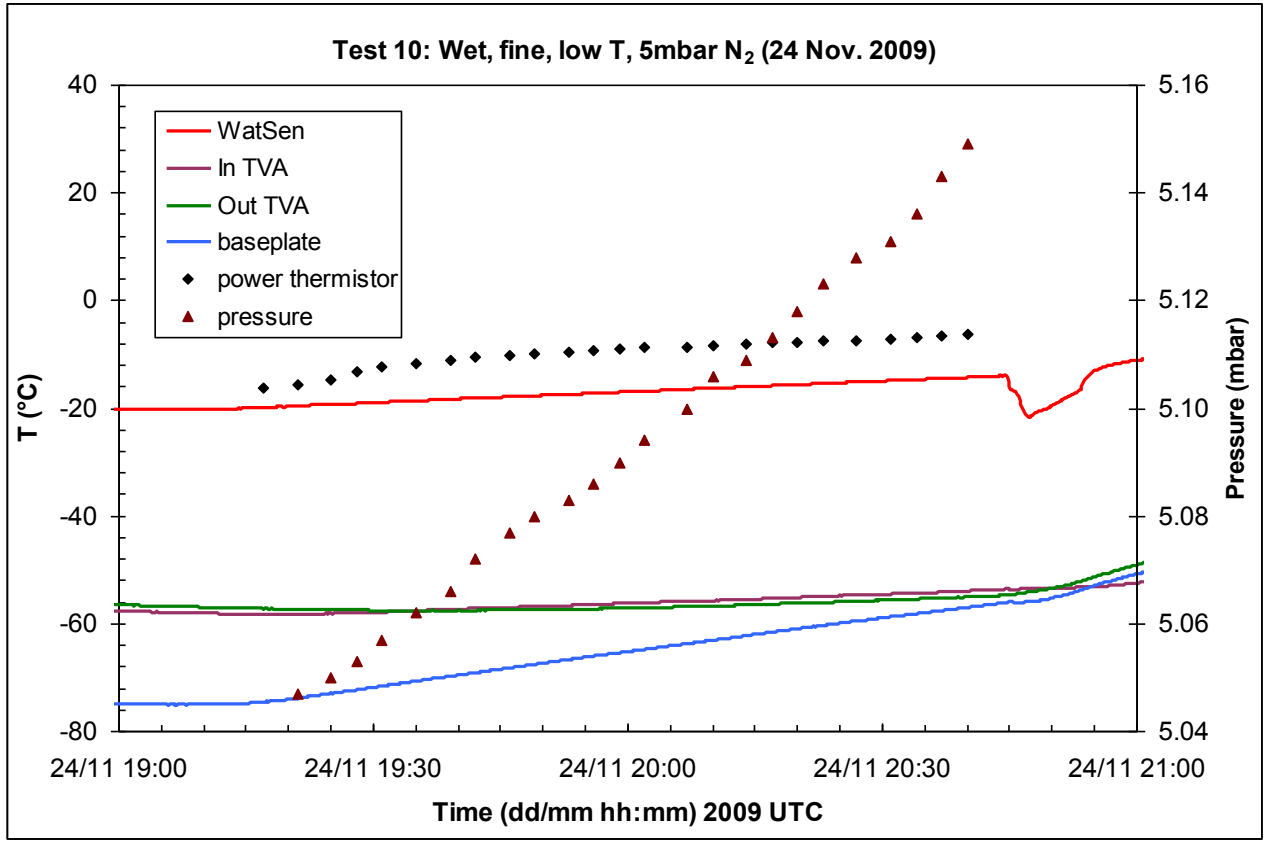

(b)

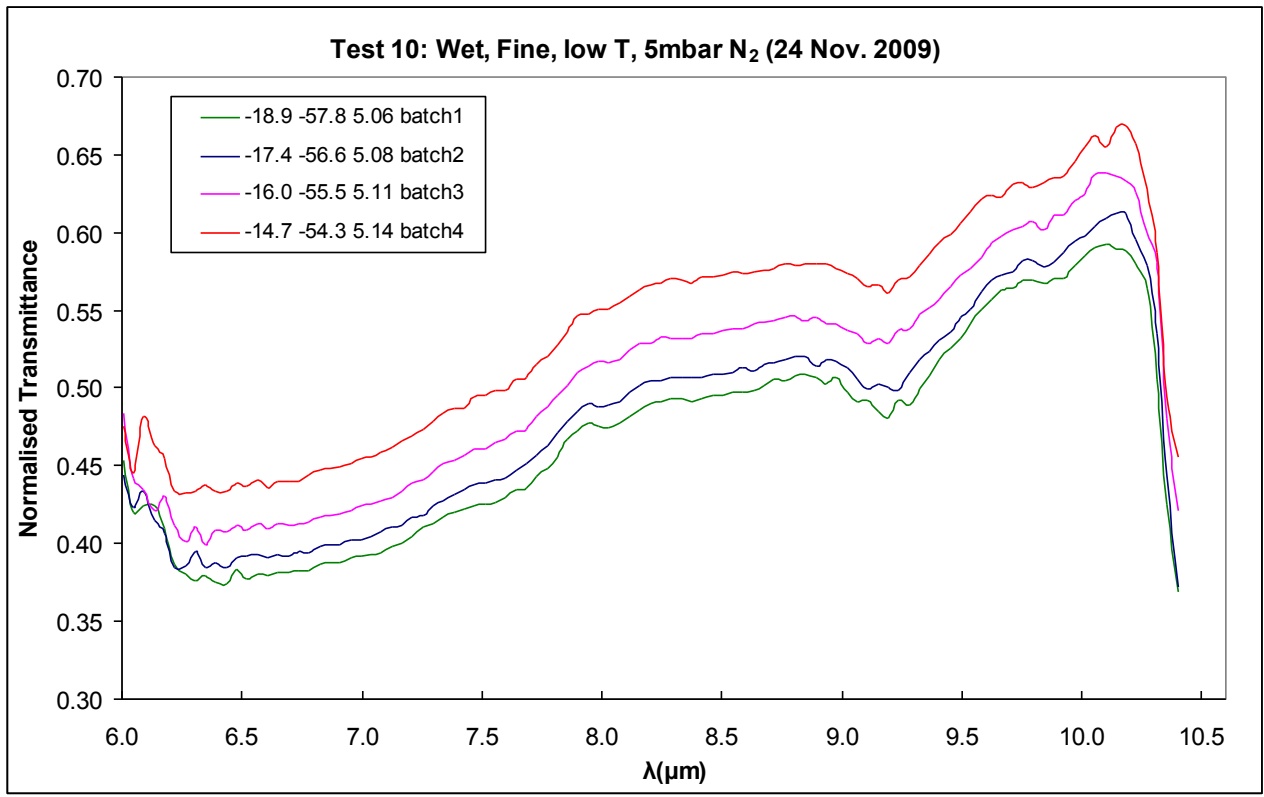

(c) (d)

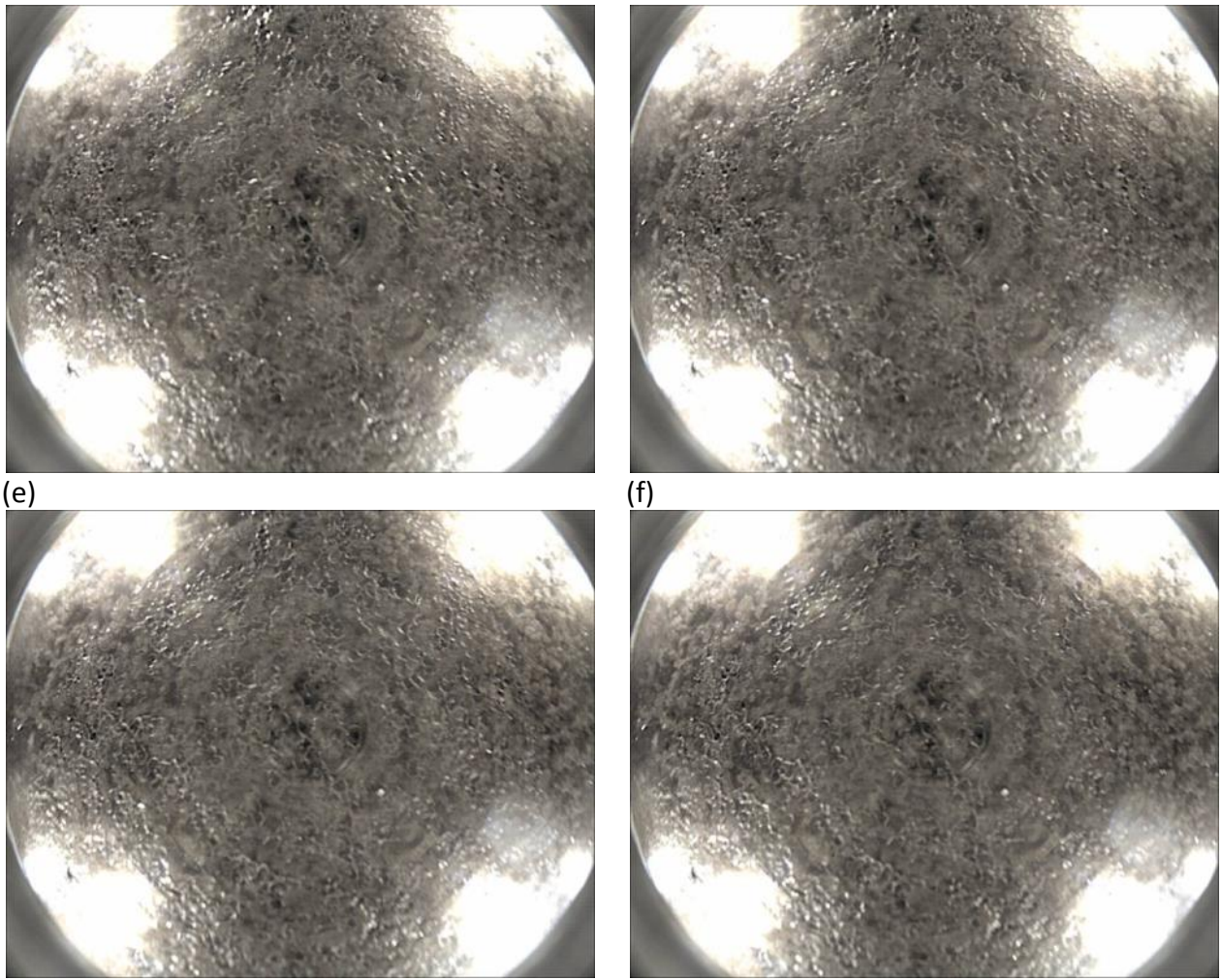

**Figure 13:** Test with wet, fine basalt, low temperature, 5 mbar N$_2$.

(a) Temperature measurements from the four environmental chamber thermistors and another associated with the internal WatSen power module. Environmental sensor placement: "WatSen" is taped to surface of WatSen just behind the diamond window; "In TVA" is taped onto cable trailing behind WatSen,"Out TVA" is taped to the surface of the TVA on the outside, at the top of the cylinder; "baseplate" is taped to the baseplate of the environment chamber;

(b) Average of each batch of 5 spectra: in the legend first number is "WatSen" thermistor (°C), second is "in TVA" soil temperature thermistor, third is pressure (mbar).

Sequence of four microscope images taken before each batch: (c) batch 1, 5.05 mbar N2, with WatSen surface temperature at – 19.6°C and with soil temperature at -58.2°C. Ice covers whole window surface, possibly pushing aside the basalt grains – this may explain why the water feature is so dominant in the spectra; (d) batch 2 (5.07mbar, -18.1°C, -57.2°C); (e) batch 3 (5.09mbar, -16.8°C, -56.1°C); (f) batch 4 (5.12mbar, -15.4°C, -54.9°C).

The final set of tests were those in which small amounts of magnesium carbonate were mixed with the basalt powder (Figure 14). The FTIR TNNS spectrum of pure magnesite has a single feature at ~7 μm, whilst hydromagnesite has a doublet at ~6.7 μm and 7.1 μm; the 7.1 μm feature is slightly stronger. When 5 wt% hydromagnesite was added to basalt, the FTIR spectrum shows a weak, but still readily identifiable doublet at the same position as in the pure hydromagnesite. For WatSen tests, when 20 wt% water was added to the mix, the resulting spectrum had a pronounced doublet, but the 6.7 μm feature is stronger. The strong carbonate feature dominates over water. It is possible that the water has reacted with hydromagnesite, and is appearing as an enhanced water-related feature in the hydromagnesite spectrum. Figure 15 shows example microscope images from this test, which give the clearest instance of water evaporating from the surface of the diamond window between batches.

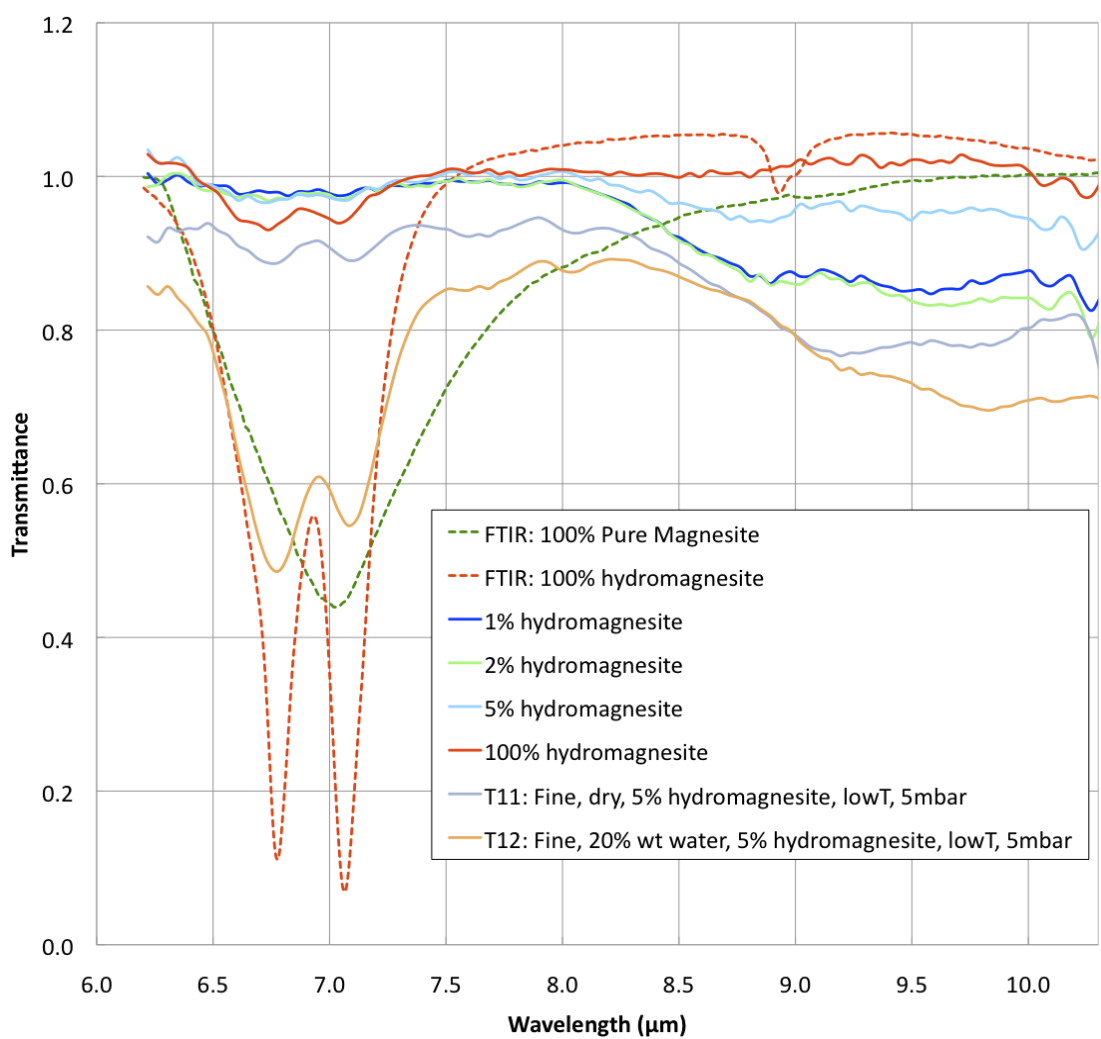

**Figure 14:** Carbonate tests. Dotted lines labelled FTIR were taken using the commercial FTIR TNNS system. T11 and T12 were taken using the TVA in the environment chamber. The spectra shown as solid curves marked 1, 2, 5 and 100 wt% hydromagnesite were acquired by simply laying the material on top of the diamond window at room temperature/pressure, rather than embedding WatSen in the TVA. In these experiments, it was more difficult to identify hydromagnesite, presumably because contact with the window was poor.

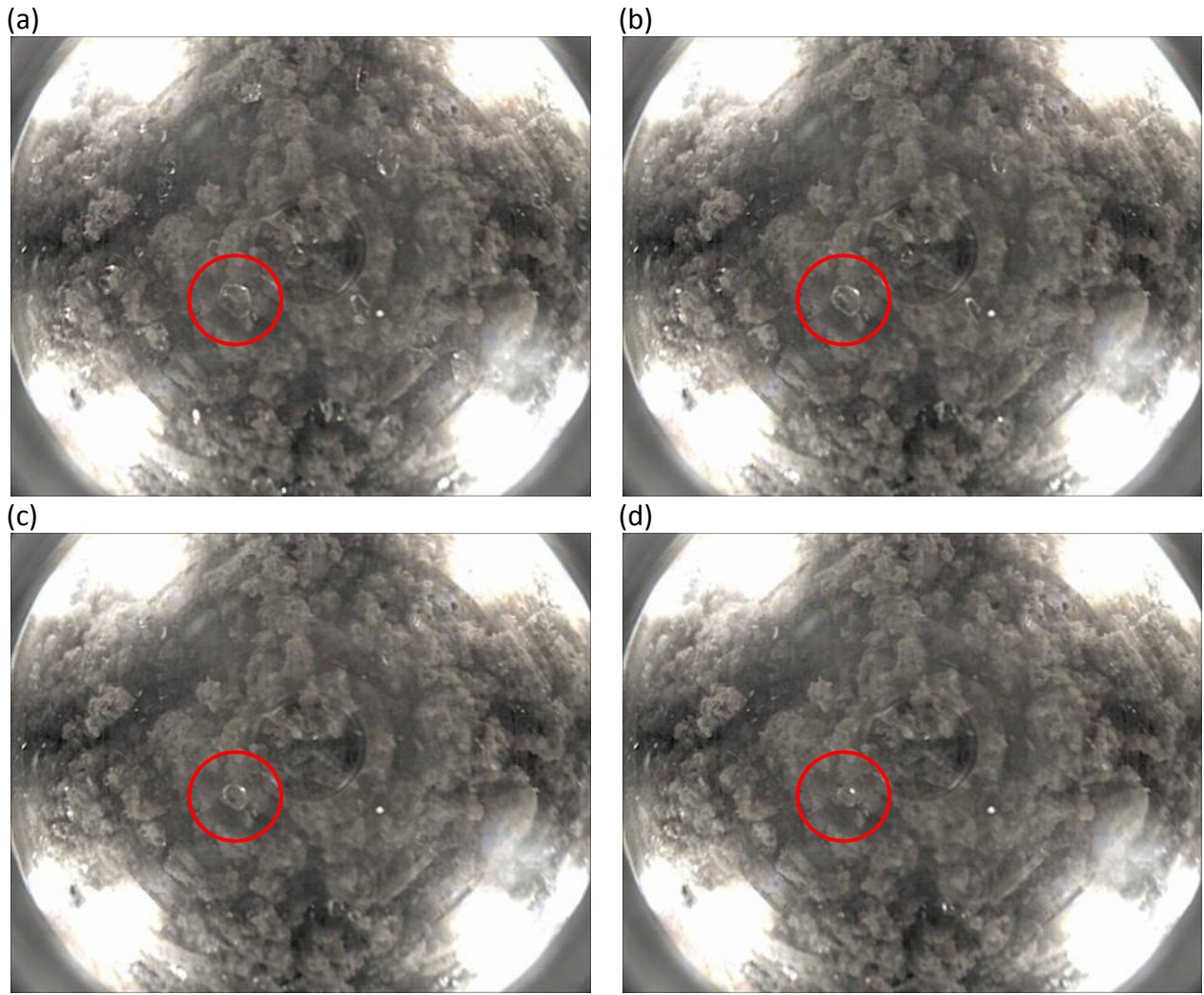

**Figure 15:** Series of four microscope images taken before each batch of spectra for test with 20 wt% water, 5 wt% magnesite, fine basalt. The red circle highlights a frozen water droplet on the ATR window. (a) WatSen surface temperature -50.47°C and pressure 5.062 mbar before batch 1 of spectra are taken. Frozen drops of water can be seen on window; (b) batch 2: -50.77°C, 5.211 mbar. Frozen drops of water seem to be smaller in size suggesting sublimation is occurring; (c) batch 3: -51.36°C and pressure 5.349 mbar; (d) batch 4: -51.77°C , 5.478 mbar. Frozen drops of water seem to be even smaller in size suggesting sublimation is continuing.

## 4. Conclusions

We have designed and built a prototype compact breadboard model of an instrument designated WatSen: a combined ATR spectrometer, fixed-focus microscope and humidity sensor that meets ESA's design specifications in terms of mass and power budget. The current TRL of the instrument is somewhere between 5 and 6 (NASA and ESA definitions, see http://en.wikipedia.org/wiki/Technology_readiness_level; accessed 21st November 2012) .

WatSen performs to specification at room temperature and pressure, and is able to record a basalt spectrum similar to that produced by a commercial ATR system. WatSen also performs to specification at the low temperatures and pressures associated with Mars' surface. WatSen is able to detect water at around the 5 wt% level in a basalt/water mix, and able to detect carbonate at the 5 wt% level in a basalt/carbonate mix. Microscope images are readily obtainable under all tested conditions, and relative humidity can also be measured.

WatSen could potentially be deployed on a variety of planetary missions in addition to Mars: for example a lunar mission to search for trace water-ice in shadowed craters, or to search for organics on the icy moons of Jupiter and Saturn, or even to the low-gravity environments of asteroids or comets. Any kind of deployment can be utilised that will put the instrument in contact with the analysed medium. For example, WatSen would be an ideal instrument to incorporate into a penetrator due to its appropriate ruggedness and slim dimensions, and its optimum performance when buried just under the surface.

**Acknowledgements**: The work was funded under contract to the European Space Agency to the Open University; ESTEC CONTRACT No 20623/07/NL/NR (WatSen microscope), to Norsk Elektro Optikk AS. We thank an anonymous reviewer for their insightful suggestions on how to improve this manuscript.